# The oxidative stress response: a proteomic view

Thierry Rabilloud[1], Mireille Chevallet[1], Sylvie Luche[1] and Emmanuelle Leize-Wagner[2,+].

[1] CEA- Laboratoire d'Immunochimie, DRDC/ICH, INSERM U 548
CEA-Grenoble, 17 rue des martyrs, F-38054 GRENOBLE CEDEX 9, France

[2] Laboratoire de Spectrométrie de Masse Bio-Organique, UMR CNRS 7509, ECPM, 25 rue Becquerel, 67008 STRASBOURG Cedex, France
+ present address : Institut de Sciences et ingéniérie Supramoléculaire, UMR CNRS 7006, Université Louis Pasteur, 8 rue Gaspard Monge, 67083 STRASBOURG CEDEX, France

Correspondence :
Thierry Rabilloud, DRDC/ICH, INSERM U 548
CEA-Grenoble, 17 rue des martyrs,
F-38054 GRENOBLE CEDEX 9
Tel (33)-4-38-78-32-12
Fax (33)-4-38-78-98-03
e-mail: Thierry.Rabilloud@ cea.fr

Abstract

The oxidative stress response is characterized by various effects on a range of biological molecules. When examined at the protein level, both expression levels and protein modifications are altered by oxidative stress. While these effects have been studied in the past by classical biochemical methods, the recent onset of proteomics methods have allowed to study the oxidative stress response on a much wider scale. The input of proteomics in the study of oxidative stress response or in the evidence of an oxidative stress component in biological phenomena is thus reviewed in this paper.

1. **Introduction**

Oxidative stress can be defined as the toxic effect of chemically reactive species derived from oxygen (superoxide, peroxide, hydroxyl radical) or mixed nitrogen-oxygen species (NO, peroxynitrite). Among those, the most active species is hydroxyl radical (OH·), most generally produced from hydroperoxides by the Fenton reaction, using reducing agents and metallic ions able of oscillating between different valencies (e.g. Cr, Mn, Fe, Ce). This toxic effect can take place on various biological molecules, such as carbohydrates, unsaturated lipids, proteins, or nucleic acids. Various damages can result on these substrates. Unsaturated lipids are transformed into lipid peroxides, a reaction which is autoamplified with molecular oxygen. Proteins undergo various damages. The peptide backbone can be cut by hydroxyl radical, and many side chains are prone to various modifications. Examples include tyrosine hydroxylation (by hydroxyl radical), or nitration (by peroxynitrite), methionine or cysteine oxidation (by hydroxyl radical or peroxide). However, one of the most common modification on protein side chains is carbonyl formation (while normal proteins do not have any keto or aldehyde groups). Depending on the nature of the amino acid side chain, this modification can arise by various mechanisms. While aliphatic side chains (e.g. Val, Leu, Ile) require the very reactive hydroxyl radical to be oxidized, more reactive side chains (e.g. Trp) can be oxidized by less active oxidants (e.g. peroxide). There is also some evidence of lysine and arginine oxidation by metal catalyzed oxidation [1, 2].

Nucleic acids are also targets for oxidative stress-induced damage. Because of the stable nature of nucleic acids, only hydroxyl radical can oxidize them. One of the most common reaction is the cleavage of the phosphosugar backbone, in a reaction quite similar to the one used for chemical DNA footprinting. The other most common reaction is guanine oxidation into 8-oxo guanine, which leads to mutations at the next replication cycle, as 8-oxoG pairs with adenine rather than cytosine.

It must be recalled, however, that this oxidative stress is common to all organisms living under aerobic conditions. As all chemical reactions, oxygen reduction under biological conditions is far from being perfect, and partial reduction with less than four electrons gives rise to reactive oxygen species, which are direct oxidative stress agents, or can combine with other molecules to yield other oxidative stress inducers (e.g. superoxide + NO gives peroxynitrite). Consequently, many organisms have evolved protection or repair mechanism as an adaptation to this constitutive oxidative stress. Examples include redox enzymes destroying the reactive oxygen species (e.g. superoxide dismutases, catalases, peroxidases, peroxiredoxins) or chemical reducers encountered in biological systems (e.g. ascorbic acid, tocopherols), or repair systems removing oxidatively-damaged components of complex molecules (e.g. lipases, nucleic acid excision systems). The redox enzymes make use of the reducing molecules produced by the general metabolism (NADH, NADPH) to provide the reducing equivalents needed to destroy the reactive oxygen species.

Besides this constitutive oxidative stress, cells can be submitted to extra bursts of oxidative stress, arising from the activity of some enzymes (e.g. oxidases), but also of some signalling pathways (e.g. TNF).

Thus, because the cells have evolved these protective systems, a real stress occurs only in situations where the oxidative power present in chemically reactive species overcomes the antioxidant defence line present in the cells.

Most of our knowledge concerning oxidative stress and the associated cellular responses has been obtained using classical biochemical approaches. Examples include the role of antioxidant enzymes as lifespan-prolonging agents in drosophila [3], or the existence of oxidative damages on proteins in oxidative stress conditions [4], using detection techniques specially designed for the detection of carbonyl groups on proteins [5]. More generally, the techniques designed to evidence the oxidative stress-induced modifications on proteins often use immunological detection of modified residues themselves (e.g. nitrotyrosine) or of haptens that can be selectively coupled to the modified residues (e.g. detection of carbonyls via immunodetection of coupled dinitrophenylhysrazine). The great sensitivity of immunodetection allows to detect minute amounts of modified proteins. This is often required, as the proteasome efficiently degrades oxidatively-damaged proteins, so that proteasome inhibitors are often used to increase the detection of oxidatively-modified proteins [6].

## 2. Proteomics studies on oxidative stress response

### 2.1. General considerations

While these dedicated biochemical techniques have proved useful, they require to test proteins one by one on an hypothesis-driven basis. Moreover, they focus only on one type of modification and do not render the complexity of the cellular response to oxidative stress. As a matter of facts, the cellular response to oxidative stress is not a passive one resulting in the bare accumulation of modified proteins. Cells submitted to oxidative stress defend themselves by various mechanisms, including metabolic switches, neosynthesis of proteins to replace the damaged ones, and active degradation of damaged proteins. Here again, classical approaches based on narrow hypotheses do not reflect the complexity of the cell response, which is a mix of changes in protein levels, controlled post-translational modifications of proteins (e.g. phosphorylation) and oxidative damages on proteins.

The onset of global approaches has considerably changed our way of investigating various complex cellular phenomena. Because oxidative stress response involves many post-translational events, proteomic approaches are much more suited to the study of this phenomenon than other global approaches such as transcriptomics.

However, not all proteomics approaches are able to deal correctly with the specific features of the oxidative stress response. Because oxidative stress is a constitutive phenomenon, mostly quantitative differences are expected. The proteomics approach chosen must therefore be able to deal with such quantitative changes. Furthermore, post-translational changes are expected to be a major component of the oxidative stress response, and the technical approaches chosen must be able to deal with complex post-translational modifications. It must be emphasized that our knowledge of post-translational modifications induced by oxidative stress is far from complete. Besides well-known controlled modifications, such as phosphorylation, the oxidative damages on proteins are not fully described, and new oxidative modifications are likely to be discovered. Furthermore, we do not have biochemical tools able to address all known oxidative modifications. For example, there are no antibodies against oxidation products of tyrosine, tryptophan or methionine.

These constraints rule out certain flavours of proteomics, such as shotgun approaches [7] or isotope-coded, peptide-based approaches [8]. As a matter of facts, most of the proteomics studies of oxidative stress response published to date have used two-dimensional electrophoresis as a protein separation and quantification tool, coupled with mass spectrometry as a protein characterization tool.

Despite its weaknesses in the separation of important classes of proteins such as membrane proteins [9],two-dimensional electrophoresis is still the most performing separation tool when dealing with complete proteins. Moreover, a rather precise quantitative analysis of relative protein amounts is possible by this technique. This technique is also able to separate proteins by rather subtle variations of the isoelectric point (pI) , and many post-translational modifications can be detected by this way. The most typical example is phosphorylation, but some stages of cysteine oxidation (cysteine sulfinic and sulfonic acids) are also expected to induce pI changes. Last but certainly not least, the separation power of two-dimensional gels considerably simplifies the subsequent analysis by mass spectrometry. As a matter of facts, a spot on a two-dimensional gels contains at most 2-3 proteins. The digestion of these will produce a few tens of peptides, a complexity that is easily dealt with by peptide mass fingerprinting approaches based on MALDI-MS or MALDI-MS/MS, or by nanoelectrospray LC/MS or LC/MS/MS approaches. This low complexity affords in turn an important sequence coverage, which means that peptides covering an important part of the complete sequence of the proteins of interest are analyzed in the mass spectrometer. This implies in turn that the likelihood of finding one of the few modified peptides in a protein altered by oxidative stress is higher with this technical setup than with other ones dealing with more complex peptide mixtures (e.g. shotgun approaches). Furthermore, a previous knowledge of the modification is not needed, and MS/MS approaches often allow both the description of the modification and the localization of the modification in the modified peptide. This statement does not mean however, that such an assignment is trivial. As a rule of thumb, most modifications considerably decrease the MS signal. Several mechanisms can take place to explain this decrease. First of all, many modifications alter the charge of the peptide and often bring negative charges (e.g. phosphorylation, sulfation, cysteine oxidation in sulfinic or sulfonic acid). This decrease in turn the ionization efficiency of peptides in cations, which is the usual and optimal ionization mode. In some extreme cases, it can be even better to take advantage of the strong negative charge and to analyze peptides in the negative ion mode [10, 11]. Other modifications (e.g. glycosylation) are so heterogeneous that they split the signal of one peptide into many small peaks. In other cases (e.g. carbonylation) the modification favors peptide-peptide interactions (e.g. by Schiff base formation), which decreases in turn the peptide extraction yields and thus the signal. For these reasons, modification assignment is not easy and poorly documented except for phosphorylation, for which dedicated strategies have been described (reviewed in [12, 13])

**2.2. Studies on prokaryots and lower eukaryots**

Examples of proteomics studies of the oxidative stress response cover various biological conditions and various organisms. Starting with the simplest organisms, i.e. bacteria, such studies cover both aspects of the oxidative stress response. One aspect is how bacteria resist to a major oxidative stress, and a good example of this type of study is provided by the analysis of proteins induced by the presence of hydrogen peroxide in the bacterium *Francisella tularensis* [14]. This is not only an in vitro model, as this bacterium survives in macrophages and is thus able to resist to the oxidant species produced by the NADPH oxidase in the macrophage phagolysosome. The proteins induced in this bacterium by oxidative stress are typical of a complex oxidative stress response. Besides antioxidant proteins (AhpC), many chaperones are also induced (to refold proteins unfolded by oxidative stress), as well as proteases (to degrade irreversibly damaged proteins and thus avoid the deleterious effects of their misfunction). Interestingly enough, the iron metabolism is also changed, in this case by the induction of a bacterioferritin. More detailed results were also obtained from a more characterized bacterium, *Bacillus subtilis*, used as a model organism [15]. Here again, there is a clear induction of antioxidant enzymes (peroxiredoxins, catalase and superoxide dismutase) and also a clear trend in iron metabolism aiming at reducing the free iron content in the cell. This can be explained by the fact that iron react with peroxide to produce hydroxyl radical, which is the most active and the most toxic oxidant species. The modulation of iron metabolism is thus an important regulation to decrease the overall toxicity of the peroxide.

Apart from these protein inductions, proteomics has also been used to search for proteins that are modified by an oxidative stress. Typical examples of such proteins are peroxiredoxins and GAPDH [16], in which the active site cysteines are oxidized into cysteic acid upon oxidative stress. As this modification induces a pI shift in the protein, it is easy to find through a 2D electrophoresis-based approach. Other, more subtle oxidative modifications occurring on cysteines, e.g. disulfide bond formation cannot be detected as easily and require dedicated approaches [17], using sequential labelling of thiols prior to or following chemical reduction. These approaches clearly show that several bacterial proteins undergo disulfide bond formation upon oxidative stress in E coli [17].

Proteomics can also be used in a different way, i.e. to demonstrate an oxidative stress component in a complex biological response, generally by the observation of an induction of antioxidant proteins. A good example is provided by the study of the stationary phase in *Bacillus licheniformis* [18]. This work clearly shows that many proteins induced by an experimental oxidative stress are also found induced during the stationary phase. This includes not only antioxidant proteins (catalase and peroxiredoxins) but also chaperones.

This concept of unicellular model organisms has been extended to eukaryots to further investigate the oxidative stress response. A very precise work has been made on the yeast Saccharomyces cerevisiae to identify changes in protein expression induced by peroxides [19]. Interestingly enough, antioxidant enzymes are not so prominent in this response, which rather emphasises a deep reorientation in the central metabolism. As an example, glycolysis is decreased while the pentose phosphate pathway is increased. While glycolysis produces NADH, the pentose phosphate pathway produces NADPH, which is the reducing substrate used by key antioxidant systems in eukaryots, the glutathione peroxidase system and the peroxiredoxin system.

**2.3. Studies on mammalian systems**

The perspective is generally fairly different when the oxidative stress response of mammalian cells is studied. As a matter of facts, the metabolism of these cells is much less flexible than the one of bacteria, yeasts or plant cells, so that metabolic adaptation is much less obvious. However, such an adaptation in the glucide metabolism has been observed in cell lines adapted to survive in low concentration of hydrogen peroxide [20]. However, The increased abundance of aldose reductase can also be interpretated as a detoxifying mechanism toward unsaturated or hydroxylated aldehydes, which are reduced by aldose reductase in the corresponding alcohols. These aldehydes are toxic end products of the degradation of lipid peroxides, while the corresponding alcohols are much less toxic. There are only a few studies of this type, examining how mammalian cells can adapt to a constitutive, increased oxidative stress. A good example is provided for myeloid [21] and epithelial cells [22] . These papers show that the metabolic adaptation of mammalian cells is much more difficult to understand than the one of bacteria or yeasts. Besides the induction of some redox regulators (peroxiredoxins in [21] and [22], superoxide dismutase in [20]) a change in only a few enzymes involved in the glycolysis pathway was observed. In addition, only a few proteins involved in protein folding are induced.

Besides these studies on the adaptation to oxidative stress, more studies focus on proteins as targets under oxidative stress conditions. One of the consequences of an oxidative stress may be the decrease in some protein content in the cell, as a consequence of the oxidative stress damage and of the induced protein degradation [23]. Here again, it is difficult to draw a clear trend from the variety of proteins that appear to be decreased upon oxidative stress.

However, most of the proteomics studies on the oxidative stress response of mammalian cells deal with the wide scale identification of target proteins for a defined modification. There are only a few that try to identify modifications without a prior hypothesis. Among those, the studies devoted to peroxiredoxins are a good example of the power of the two-dimensional electrophoresis/ mass spectrometry approach. The starting point of the study is the simple observation of an altered migration of some proteins after oxidative stress of the cell. MS analysis first identifies these proteins as the various members of the peroxiredoxin family, and then shows that the modification is indeed the overoxidation of the active site cysteine of these proteins [24]. However, this study also shows that such an approach is not as straightforward as it may seem. The oxidation of cysteine into cysteine sulfinic or cysteic acid brings a strong negative charge on the modified peptide, thereby greatly reducing the ionization of the modified peptide in the standard cationic mode used for peptide mass spectrometry. Consequently, the detailed study of this phenomenon, which has shown that this modification is reversible in cells [11, 25], has required the development of dedicated mass spectrometry techniques [10]. Interestingly enough, this cysteine modification is not limited to peroxiredoxins, and has also been observed in other proteins [26]. This latter work has also demonstrated other migration alterations upon oxidative stress, which remain to be characterized, as well as quantitative changes. Moreover, the evidence of the reversibility of this modification [11], [25], has prompted active research to identify the enzymatic systems responsible for this regeneration [27, 28].

As the thiol group of cysteine is one of the most sensitive redox groups in proteins, other proteomics studies have aimed at the identification of other thiol modifications upon oxidative stress. The formation of disulfide bridges has been investigated using off-diagonal electrophoresis [29], and this resulted in the evidence of complex interprotein bridges. Other studies have aimed at the identification of disulfide bridges brought upon oxidative stress between protein thiols and glutathione [30]. Last but not least, a simple but elegant identification methods of free thiol groups in mitochondria has been described [31]. This method has been recently applied to the identification of oxidized proteins induced upon alcohol consumption [32], a topic which has also been addressed by other methods involving differential thiol labelling [33].

However, in all these cases the variety of proteins involved does not allow an easy interpretation of the results.

Two other common, oxidative stress-induced protein modifications are also frequently assessed in proteomics studies, namely tyrosine nitration and carbonyl formation. This is due to the availability of antibodies against the nitrophenyl and dinitrophenyl haptens, which enable a sensitive and specific detection of these modifications.

These modifications are looked for in two type of studies. In the first type, the cell or tissue of interest is submitted to a well-characterized oxidative stress, and the target proteins for this stress are investigated using the modification specific assay by blotting and the resolving power of proteomics. Examples of such studies have been carried out on muscle [34] or on cells treated with a well-known oxidative stress inducing toxin [35], and have led to the identification of a few target proteins. However, precise assignment of the modification sites has not been carried out in these studies.

**3. Cell biology proteomics studies related to oxidative stress**

Another use of the molecular tools detecting oxidative modifications is to assay these modifications in biological situations where an oxidative component is suspected. By this way, an increase in the amount of modified proteins is a hallmark of an oxidative stress. Furthermore, the resolving power of proteomics allows to identify the target proteins of this oxidative proteins. Such approaches have been used for tyrosine nitration, e.g. in aging [36], for protein carbonyl formation, e.g. in Alzheimer disease [37, 38], or in other neurodegenerative disorders [39] or in aging [40, 41] or in apoptosis [42] and for other oxidative stress-related modifications, such as conjugation with 4 hydroxynonenal [43], which is a by-product of the oxidative degradation of unsaturated fatty acids

While the expression data are usually difficult to interpret, the data obtained by the study of protein carbonyls are much simpler, and show an obvious trend for the damage of various chaperone proteins [37], [42], including mitochondrial chaperones. A review of mitochondrial changes induced by oxidative stress and detected by proteomics approaches has been recently published [44]. Most often, the proteomics studies aiming at the identification of oxidatively modified proteins also take into account the increase or decrease in protein amounts (e.g. [40,42]). However, some studies focus only on protein amount changes, regardless of any modification-specific study, in biological situations where oxidative stress is suspected. In this case, it is the changes in antioxidant enzymes which becomes the hallmark of oxidative stress. Very different biological situations have been investigated by this approach, including arsenite toxicity [45], the effect of diesel exhaust particles [46], neurodegeneration [47], brain ischemia [48], chemically-induced cirrhosis [49], aging in liver [50] or growth factor signalling [51]. The changes are generally detected for peroxiredoxins and glutathione peroxidases, but sometimes for other proteins such as heme oxygenase [45, 46].

## 4. Expert opinion

The most important weakness in the proteomics studies of the oxidative stress response is clearly linked with the limited performance of the current proteomics techniques. Apart from the fact that membrane proteins are strongly underrepresented [9], which is clearly a problem, especially when trying to find out target proteins for oxidative stress [44], the representation of the real proteome may be very variable from one system to another. As a matter of facts, the resolving power of two-dimensional gels is 1000-2000 protein spots. When taking into account the chemical diversity brought by post-translational modifications, this resolving power is close to 50% of the total proteome, and can reach up to 70% when multiple gels are used [52]. However, it drop down to ca. 10% in mammalian cells, where many more genes are expressed and where the level of post-translational modifications is much higher than in prokaryots. This means in turn that our proteomics view of mammalian cells is very limited to the more soluble and more abundant proteins. This may explain why proteomics results obtained in oxidative stress response are often so difficult to interpret, as they only provide a very partial view of the whole picture.

Alternate methods such as shotgun proteomics, based on multidimensional chromatographic separation of peptides [7], does not present the same drawbacks and is able to analyze many more peptides with a much less bias against membrane proteins. However, this approach is very difficult to make quantitative and offers limited sequence coverage when mixtures as complex as whole cell extracts are analyzed. This makes these methods less suited for analysis of oxidative stress, which is mostly a quantitative phenomenon in which post-translational modifications play important role, which makes the sequence coverage a crucial parameter.

Apart from this classical representation problem in proteomics, which are a bottleneck in all proteomics studies, another more specific problem arises in the study of the oxidative stress. Many interesting modifications on target proteins arise from oxidative damages, which can take very different forms. One of the most documented modifications is carbonyl formation. This modification is usually detected by conjugation of the carbonyl group with an hydrazine or hydrazide, and the resulting hydrazone is assayed by detection of the chemical moiety coupled with the hydrazine group. This can be by an antibody-hapten reaction, or by avidin-biotin affinity [53]. However, this process is far from being perfect, as the hydrazone formation is very pH-dependent and easily decreased by steric or electrostatic interactions. Furthermore, the excess reactant must be removed by precipitation of proteins. This process induces protein losses and also partial cleavage of the hydrazone.

While other modifications such as tyrosine nitration are efficiently detected by the way of specific antibodies, there are however many other modifications for which no molecular tools are available. This is for example the case of tyrosine hydroxylation, but also of methionine or tryptophan oxidation. It is also very likely that some modifications remain to be discovered.

In this scenario, the pI shift that is often associated with oxidative modifications may be a good tool to separate the modified forms from the bulk of unmodified proteins. The resolving power of narrow-range pH range should improve the number of cases in which a modified protein will be separated from the unmodified one. It is then crucial to achieve an optimal sequence coverage in mass spectrometry to maximize the probability of identifying the modification and assigning it as precisely as possible ideally to a single amino-acid (e.g. in [24]). This is clearly an important area of improvement over most of the studies published to date, where the modifications are most often just assigned to a protein and not more precisely than that.

**5. Five-year view**

The expectations that ca be put five years ahead lie on the directions outlined in the previous paragraph. On the one hand, the continuous progress in protein separations will increase our representation of proteomes in general, and this will of course have an impact in the studies of oxidative stress. This also holds true for the mass spectrometry side, which becomes always more precise and comprehensive. A quantum leap in the precision of the description of the oxidative stress-induced modifications is very likely in the next few years, aiming at assigning in most cases the precise site of modification. Together with improved structural knowledge or predictions in protein structure, this will considerably help in understanding how modifications modulate protein function. Her again, peroxiredoxins provide a very good example, as the normal on oxidized form show neither the same structure nor the same function [54].

The situation of the molecular tools, and especially antibodies, for detecting modifications is however much less encouraging, and it is very likely that the modifications escaping detection today will still not be detected by this means within the next few years.

However, one recently-introduced technique should be very useful for the study of oxidative stress-induced modifications. This technique has been nicknamed COFRADIC, for Combined FRACtions DIagonal Chromatography [55]. The rationale of the technique is to select peptides in a complex mixture on the basis of a chemical property. To this purpose, the complex peptide mixture is first fractionated by reverse phase chromatography, and fractions are collected. All fractions are then submitted to a chemical treatment that will specifically modify some peptides in the mixture. Each fraction is then refractionated on the same column than in the first fractionation. The bulk of unmodified peptide will elute at the same position in the gradient and these are discarded. The modified peptides will elute either earlier or later in the gradient, depending if the chemical treatment makes them more hydrophilic or more hydrophobic. Those peptides are then selectively analyzed in the mass spectrometer. This approach has been recently reviewed and described in detail [56]. The great interest of this method is its versatility, and many different modifications schemes can be devised to detect classes of peptides. Interestingly enough, the proof of concept of the method has been made on methionine-containing peptides, and the chosen modification was methionine oxidation with hydrogen peroxide [55]. This means conversely that peptides containing oxidized methionine can be selectively analyzed by making a reduction between the two separations, e.g. with borohydride. Carbonyl-containing peptides should also be selectable, e.g. with reductive amination with cyanoborohydride, or hydrazone formation with a less bulky hydrazine than the usual dinitrophenylhydrazine.

The versatility of organic chemistry should greatly help in designing modification schemes able to select many of the oxidative modifications that are today difficult to analyze (e.g. tyrosine or tryptophan oxidation, but also oxidation at other amino acids). Furthermore, the concept can be increased to an even greater flexibility by investigating the usefulness of peptide cleavages. Up to now, rather subtle modifications have been used. They do not modify too much the overall properties of the peptides, so that some fractions can be pooled in the second analysis (hence the name COmbined FRActions). For example, fractions 1, 11, 21 etc, can be analyzed in a single run, as it is very unlikely that a methionine oxidation or a phosphate removal will move one peptide as wide as ten fractions. However, if the experimentators are prepared to run fractions one by one, then a peptide cleavage, which will induce dramatic changes in the chromatographic mobility between the parent peptides and the cleavage peptides, can be used. As to oxidative stress, this could be very useful to analyze oxidation of aromatic amino acids. As a matter of fact, many :classical methods for chemical cleavage after aromatic amino acids (Tyr or Trp), start with the oxidation of the amino acids, followed by a cleavage step that uses the reactivity of the oxidized amino acid [57, 58]. The use of the sole cleavage step could provide an efficient tool to localize oxidized aromatic amino acids.

As for all peptide-based proteomics approaches, it is difficult to perform a quantitative analysis with this method. However, it is possible to devise a two-stage approach allowing a relative quantification of the modified peptide. The first stage consists in the determination of the modified peptide with the COFRADIC approach. Once the peptide has been identified and characterized, it is possible to predict via sequence analysis the mass of the unmodified peptide. In the second stage, a standard LC/MS analysis then allows to reach the intensity ratio between the MS signals for the modified vs. unmodified peptide. If the modification is discrete enough, it should minimally alter the ionization efficiency for the modified peptide compared to the normal one, so that the signal ratio between the two is a good approximation of the real quantitative ratio.

In conclusion, the coming years should demonstrate a major improvement in the description of the oxidative stress-induced modifications of proteins, both in the variety of modifications assessed and in the precision of the determination of the modified sites. This should provide much more precise insights in the deleterious effects of oxidative stress on proteins. As a matter of facts, it should be kept in mind that the data obtained by proteomics on protein modifications are just a first step. The following one is to demonstrate that the modifications observed on a protein alter its activity in a way or another, either directly by altering its conformation and/or the active site (e.g. cysteine oxidation in the active site of peroxiredoxins or GAPDH), or by altering the association to other proteins. Such rigorous demonstrations should also considerably enhance our comprehension of the deleterious phenomena occurring during oxidative stress

## 6. Key issues

-Proteomic studies allow to study the oxidative stress response with a much wider view than conventional biochemical methods.

-Proteomic studies also allow to evidence the oxidative stress component in various complex cellular processes, including cell signalling, aging and pathologies

-With the development of dedicated biochemical tools and their adaptation in the proteomics toolbox, it is now possible to study both protein expression variations and oxidative protein modifications (e.g. tyrosine nitration or carbonyl formation) by the wide-screen proteomics toolbox

-The current proteomics toolbox is not comprehensive enough, however, to deliver a clear picture of the complex cellular events taking place upon oxidative stress. The data obtained to date are generally very fragmented.

-Despite this partial character, the importance of subclasses of antioxidant enzymes (e.g. peroxiredoxins) and of chaperone proteins, has been exemplified in numerous studies of the oxidative stress response.

-Besides the protein analysis scope, which is a major challenge in proteomics and which will be beneficial to all proteomics studies, the study of the oxidative stress response would benefit of dedicated improvements, especially in the detection of nowadays poorly characterized oxidative modifications (e.g. tyrosine hydroxylation)

# References

*[1] Requena JR, Chao CC, Levine RL, Stadtman ER. Glutamic and aminoadipic semialdehydes are the main carbonyl products of metal-catalyzed oxidation of proteins. *Proc Natl Acad Sci U S A*. 98(1):69-74. (2001).
Indispensable to understand oxidative protein carbonyl formation

[2] Requena JR, Stadtman ER. Conversion of lysine to N(epsilon)-(carboxymethyl)lysine increases susceptibility of proteins to metal-catalyzed oxidation. *Biochem Biophys Res Commun.* 264(1):207-211 (1999).

[3] Orr WC, Mockett RJ, Benes JJ, Sohal RS. Effects of overexpression of copper-zinc and manganese superoxide dismutases, catalase, and thioredoxin reductase genes on longevity in Drosophila melanogaster. *J Biol Chem.* 278(29):26418-26422 (2003).

[4] Yan LJ, Levine RL, Sohal RS. Oxidative damage during aging targets mitochondrial aconitase. *Proc Natl Acad Sci U S A*. 94(21):11168-11172 (1997).

[5] Levine RL, Williams JA, Stadtman ER, Shacter E. Carbonyl assays for determination of oxidatively modified proteins. *Methods Enzymol.* 233:346-357 (1994)

[6] Drake SK, Bourdon E, Wehr NB, Levine RL, Backlund PS, Yergey AL, Rouault TA. Numerous proteins in Mammalian cells are prone to iron-dependent oxidation and proteasomal degradation. *Dev Neurosci.* 24(2-3):114-124 (2002).

[7] Washburn MP, Wolters D, Yates JR 3rd. Large-scale analysis of the yeast proteome by multidimensional protein identification technology. *Nat Biotechnol*. 19(3):242-247 (2001).

[8] Gygi SP, Rist B, Gerber SA, Turecek F, Gelb MH, Aebersold R. Quantitative analysis of complex protein mixtures using isotope-coded affinity tags. *Nat Biotechnol.* 17(10):994-999 (1999).

[9] Santoni V, Molloy M, Rabilloud T. Membrane proteins and proteomics: un amour impossible? *Electrophoresis.* ;21(6):1054-1070 (2000)

[10] Wagner E, Luche S, Penna L, Chevallet M, Van Dorsselaer A, Leize-Wagner E, Rabilloud T. A method for detection of overoxidation of cysteines: peroxiredoxins are oxidized in vivo at the active-site cysteine during oxidative stress. *Biochem J.* 366(Pt 3):777-785 (2002).

[11] Chevallet M, Wagner E, Luche S, van Dorsselaer A, Leize-Wagner E, Rabilloud T. Regeneration of peroxiredoxins during recovery after oxidative stress: only some overoxidized peroxiredoxins can be reduced during recovery after oxidative stress. *J Biol Chem.* 278(39):37146-37153 (2003).

[12] Reinders J, Sickmann A.
State-of-the-art in phosphoproteomics.
*Proteomics.* (2005) in press

[13] Areces LB, Matafora V, Bachi A.
Analysis of protein phosphorylation by mass spectrometry.
*Eur J Mass Spectrom*;10:383-392 (2004).

[14] Lenco J, Pavkova I, Hubalek M, Stulik J.Insights into the oxidative stress response in Francisella tularensis LVS and its mutant DeltaiglC1+2 by proteomics analysis.
*FEMS Microbiol Lett.* 246(1):47-54 (2005)

*[15] Mostertz J, Scharf C, Hecker M, Homuth G. Transcriptome and proteome analysis of Bacillus subtilis gene expression in response to superoxide and peroxide stress.
*Microbiology.* 150(Pt 2):497-512 (2004).
A typical study on a prokaryot, examplifying the analysis detail

[16] Weber H, Engelmann S, Becher D, Hecker M.
Oxidative stress triggers thiol oxidation in the glyceraldehyde-3-phosphate dehydrogenase of Staphylococcus aureus.
*Mol Microbiol.* 52:133-140. (2004)

[17] Leichert LI, Jakob U.
Protein thiol modifications visualized in vivo.
PLoS Biol. 2:e333 (2004).

[18] Voigt B, Schweder T, Becher D, Ehrenreich A, Gottschalk G, Feesche J, Maurer KH, Hecker M.A proteomic view of cell physiology of Bacillus licheniformis. *Proteomics.* 4(5):1465-1490 (2004).

*[19] Godon C, Lagniel G, Lee J, Buhler JM, Kieffer S, Perrot M, Boucherie H, Toledano MB, Labarre J. The $H_2O_2$ stimulon in Saccharomyces cerevisiae.
*J Biol Chem.* 273(35):22480-22489 (1998).
A pioneer oxidative stress sudy by proteomics, showing the complex metabolism reorientation

[20] Keightley JA, Shang L, Kinter M. Proteomic analysis of oxidative stress-resistant cells: a specific role for aldose reductase overexpression in cytoprotection. *Mol Cell Proteomics.*

3(2):167-175 (2004).

[21] Seong JK, Kim do K, Choi KH, Oh SH, Kim KS, Lee SS, Um HD. Proteomic analysis of the cellular proteins induced by adaptive concentrations of hydrogen peroxide in human U937 cells. *Exp Mol Med*. 34(5):374-378 (2002).

[22] Chan HL, Gharbi S, Gaffney PR, Cramer R, Waterfield MD, Timms JF.
Proteomic analysis of redox- and ErbB2-dependent changes in mammary luminal epithelial cells using cysteine- and lysine-labelling two-dimensional difference gel electrophoresis. Proteomics. 5 : 2908-2926 (2005)

[23] Vorum H, Ostergaard M, Hensechke P, Enghild JJ, Riazati M, Rice GE. Proteomic analysis of hyperoxia-induced responses in the human choriocarcinoma cell line JEG-3. *Proteomics*. 4(3):861-867 (2004).

*[24] Rabilloud T, Heller M, Gasnier F, Luche S, Rey C, Aebersold R, Benahmed M, Louisot P, Lunardi J. Proteomics analysis of cellular response to oxidative stress. Evidence for in vivo overoxidation of peroxiredoxins at their active site.
*J Biol Chem.* 277:19396-19401(2002).
How to detect a new oxidative modification with the proteomics toolbox

[25] Woo HA, Chae HZ, Hwang SC, Yang KS, Kang SW, Kim K, Rhee SG. Reversing the inactivation of peroxiredoxins caused by cysteine sulfinic acid formation.*Science*. 300(5619):653-656 (2003).

[26] Paron I, D'Elia A, D'Ambrosio C, Scaloni A, D'Aurizio F, Prescott A, Damante G, Tell G. A proteomic approach to identify early molecular targets of oxidative stress in human epithelial lens cells. *Biochem J.* 378(Pt 3):929-937 (2004)

[27]  Biteau B, Labarre J, Toledano MB. ATP-dependent reduction of cysteine-sulphinic acid by S. cerevisiae sulphiredoxin. *Nature*. 425(6961):980-984 (2003).

[28] Budanov AV, Sablina AA, Feinstein E, Koonin EV, Chumakov PM. Regeneration of peroxiredoxins by p53-regulated sestrins, homologs of bacterial AhpD.*Science*. 304(5670):596-600 (2004).

[29] Brennan JP, Wait R, Begum S, Bell JR, Dunn MJ, Eaton P. Detection and mapping of widespread intermolecular protein disulfide formation during cardiac oxidative stress using proteomics with diagonal electrophoresis. *J Biol Chem.* 279(40):41352-41360 (2004)

*[30] Fratelli M, Demol H, Puype M, Casagrande S, Eberini I, Salmona M, Bonetto V, Mengozzi M, Duffieux F, Miclet E, Bachi A, Vandekerckhove J, Gianazza E, Ghezzi P. Identification by redox proteomics of glutathionylated proteins in oxidatively stressed human T lymphocytes. *Proc Natl Acad Sci U S A*. 99(6):3505-3510 (2002).
How to detect a new oxidative modification by the proteomics toolbox

[31] Lin TK, Hughes G, Muratovska A, Blaikie FH, Brookes PS, Darley-Usmar V, Smith RA, Murphy MP. Specific modification of mitochondrial protein thiols in response to oxidative stress: aproteomics approach. *J Biol Chem.* 277(19):17048-17056 (2002).

[32] Venkatraman A, Landar A, Davis AJ, Ulasova E, Page G, Murphy MP, Darley-Usmar V, Bailey SM.
Oxidative modification of hepatic mitochondria protein thiols: effect of chronic alcohol consumption.
*Am J Physiol Gastrointest Liver Physiol.* 286:G521-G527 (2004)

[33] Suh SK, Hood BL, Kim BJ, Conrads TP, Veenstra TD, Song BJ.
Identification of oxidized mitochondrial proteins in alcohol-exposed human hepatoma cells and mouse liver.
*Proteomics.* 4:3401-3412. (2004)

[34] Stagsted J, Bendixen E, Andersen HJ. Identification of specific oxidatively modified proteins in chicken muscles using a combined immunologic and proteomic approach. *J Agric Food Chem.* 52(12):3967-3974 (2004).

[35] Pocernich CB, Poon HF, Boyd-Kimball D, Lynn BC, Nath A, Klein JB, Butterfield DA. Proteomic analysis of oxidatively modified proteins induced by the mitochondrial toxin 3-nitropropionic acid in human astrocytes expressing the HIV protein tat.
*Brain Res Mol Brain Res.* 18;133(2):299-306 (2005).

[36] Kanski J, Hong SJ, Schoneich C.
Proteomic analysis of protein nitration in aging skeletal muscle and identification of nitrotyrosine-containing sequences in vivo by nanoelectrospray ionization tandem mass spectrometry. *J Biol Chem.*280(25):24261-24266 (2005)

*[37] Korolainen MA, Goldsteins G, Alafuzoff I, Koistinaho J, Pirttila T.
Proteomic analysis of protein oxidation in Alzheimer's disease brain.
*Electrophoresis.* 23(19):3428-3433 (2002).
the oxidative stress component of a pathology

[38]Castegna A, Thongboonkerd V, Klein JB, Lynn B, Markesbery WR, Butterfield DA.
Proteomic identification of nitrated proteins in Alzheimer's disease brain.
J Neurochem. 2003 Jun;85(6):1394-401.

[39] Poon HF, Hensley K, Thongboonkerd V, Merchant ML, Lynn BC, Pierce WM, Klein JB, Calabrese V, Butterfield DA.
Redox proteomics analysis of oxidatively modified proteins in G93A-SOD1 transgenic mice--a model of familial amyotrophic lateral sclerosis.
*Free Radic Biol Med.* 39:453-462 (2005)

[40] Poon HF, Vaishnav RA, Getchell TV, Getchell ML, Butterfield DA.
Quantitative proteomics analysis of differential protein expression and oxidative modification of specific proteins in the brains of old mice. Neurobiol Aging. (2005) in press

[41] Poon HF, Castegna A, Farr SA, Thongboonkerd V, Lynn BC, Banks WA, Morley JE, Klein JB, Butterfield DA. Quantitative proteomics analysis of specific protein expression and oxidative modification in aged senescence-accelerated-prone 8 mice brain. *Neuroscien*ce. 126(4):915-926 (2004)

*[42] Magi B, Ettorre A, Liberatori S, Bini L, Andreassi M, Frosali S, Neri P, Pallini V, Di Stefano A.Selectivity of protein carbonylation in the apoptotic response to oxidative stress associated with photodynamic therapy: a cell biochemical and proteomic investigation. *Cell Death Differ.* 11(8):842-852 (2004).
A typical integrated proteomics study of oxidative stess in mammalian cells

[43] Perluigi M, Fai Poon H, Hensley K, Pierce WM, Klein JB, Calabrese V, De Marco C, Butterfield DA.
Proteomic analysis of 4-hydroxy-2-nonenal-modified proteins in G93A-SOD1 transgenic mice--a model of familial amyotrophic lateral sclerosis.
*Free Radic Biol Med.* 38:960-968 (2005)

[44] Bailey SM, Landar A, Darley-Usmar V.Mitochondrial proteomics in free radical research. *Free Radic Biol Med.* 38(2):175-188. (2005).

[45] Lau AT, He QY, Chiu JF.A proteome analysis of the arsenite response in cultured lung cells: evidence for in vitrooxidative stress-induced apoptosis. *Biochem J.* 382(Pt 2):641-650 (2004).

[46] Xiao GG, Wang M, Li N, Loo JA, Nel AE. Use of proteomics to demonstrate a hierarchical oxidative stress response to diesel exhaust particle chemicals in a macrophage cell line. *J. Biol. Chem.* 50781-50790 (2003).

[47] Krapfenbauer K, Engidawork E, Cairns N, Fountoulakis M, Lubec G.
Aberrant expression of peroxiredoxin subtypes in neurodegenerative disorders.
*Brain Res.* 967(1-2):152-160 (2003).

[48] Schrattenholz A, Wozny W, Klemm M, Schroer K, Stegmann W, Cahill MA.
Differential and quantitative molecular analysis of ischemia complexity reduction by isotopic labeling of proteins using a neural embryonic stem cell model. *J Neurol Sci.* 15;229-230:261-267 (2005).

[49] Low TY, Leow CK, Salto-Tellez M, Chung MC. A proteomic analysis of thioacetamide-induced hepatotoxicity and cirrhosis in rat livers.
*Proteomics.* 4(12):3960-3974 (2004).

[50] Cho YM, Bae SH, Choi BK, Cho SY, Song CW, Yoo JK, Paik YK. Differential expression of the liver proteome in senescence accelerated mice. *Proteomics.* 3(10):1883-1894 (2003).

[51] Imamura T, Kanai F, Kawakami T, Amarsanaa J, Ijichi H, Hoshida Y, Tanaka Y, Ikenoue T, Tateishi K, Kawabe T, Arakawa Y, Miyagishi M, Taira K, Yokosuka O, Omata M. Proteomic analysis of the TGF-beta signaling pathway in pancreatic carcinoma cells using stable RNA interference to silence Smad4 expression. *Biochem Biophys Res Commun.* 318(1):289-296 (2004)

[52] Tonella L, Hoogland C, Binz PA, Appel RD, Hochstrasser DF, Sanchez JC. New perspectives in the Escherichia coli proteome investigation. *Proteomics.* 1(3):409-423 (2001).

[53] Soreghan BA, Yang F, Thomas SN, Hsu J, Yang AJ. High-throughput proteomic-based identification of oxidatively induced protein carbonylation in mouse brain. Pharm Res. 20:1713-1720 (2003)

[54] Jang HH, Lee KO, Chi YH, Jung BG, Park SK, Park JH, Lee JR, Lee SS, Moon JC, Yun JW, Choi YO, Kim WY, Kang JS, Cheong GW, Yun DJ, Rhee SG, Cho MJ, Lee SY. Two enzymes in one; two yeast peroxiredoxins display oxidative stress-dependent switching from a peroxidase to a molecular chaperone function. *Cell.* 117(5):625-635 (2004).

*[55] Gevaert K, Van Damme J, Goethals M, Thomas GR, Hoorelbeke B, Demol H, Martens L, Puype M, Staes A, Vandekerckhove J. Chromatographic isolation of methionine-containing peptides for gel-free proteome analysis: identification of more than 800 Escherichia coli proteins. *Mol Cell Proteomics.* 1(11):896-903 (2002).
*the potentiality and versatility of the COFRADIC approach

*[56] Gevaert K, Damme PV, Martens L, Vandekerckhove J. Diagonal reverse-phase chromatography applications in peptide-centric proteomics: Ahead of catalogue-omics? *Anal Biochem.* 345:18-29 (2005)
*the potentiality and versatility of the COFRADIC approach

[57] Huang HV, Bond MW, Hunkapiller MW, Hood LE. Cleavage at tryptophanyl residues with dimethyl sulfoxide-hydrochloric acid and cyanogen bromide. Methods Enzymol. 91:318-324. (1983)

[58] Savige WE, Fontana A. Cleavage of the tryptophanyl peptide bond by dimethyl sulfoxide-hydrobromic acid. Methods Enzymol. 47:459-469 (1977)